\documentclass[aps,twocolumn,showpacs,twoside,a4paper]{revtex4}
\usepackage{amsmath,amssymb,epsfig}

\def\vc#1{{\bf{#1}}}
\def\tr{\operatorname{Tr}}
\def\tbeta{\tilde{\beta}}
\def\bra#1{\left\langle{#1}\right|}
\def\ket#1{\left|{#1}\right\rangle}
\def\bracket#1#2{\left\langle{#1}|{#2}\right\rangle}
\def\conf{{\cal C}}
\def\avg#1{\left\langle{#1}\right\rangle}
\def\sign{\operatorname{sign}}
\def\sh{\operatorname{sh}}
\def\ch{\operatorname{ch}}

\begin{document}

\title{Thermodynamic and diamagnetic properties of weakly doped
antiferromagnets}

\author{Darko Veberi\v c}
\email{darko.veberic@ijs.si}
\affiliation{Jo\v zef Stefan Institute, SI-1000 Ljubljana, Slovenia}

\author{Peter Prelov\v sek}
\email{peter.prelovsek@ijs.si}
\affiliation{Jo\v zef Stefan Institute, SI-1000 Ljubljana, Slovenia and
Faculty of Mathematics and Physics, University of Ljubljana, SI-1000
Ljubljana, Slovenia}

\author{Hans Gerd Evertz}
\email{evertz@itp.tu-graz.ac.at}
\affiliation{Institute for Theoretical Physics, Technical University 
Graz, 8010 Graz, Austria}

\date{\today}

\begin{abstract}
Finite-temperature properties of weakly doped antiferromagnets as
modeled by the two-dimensional $t$-$J$ model and relevant to
underdoped cuprates are investigated by numerical studies of small
model systems at low doping. Two numerical methods are used: the
worldline quantum Monte Carlo method with a loop cluster algorithm and
the finite-temperature Lanczos method, yielding consistent results.
Thermodynamic quantities: specific heat, entropy and spin
susceptibility reveal a sizeable perturbation induced by holes
introduced into a magnetic insulator, as well as a pronounced
temperature dependence. The diamagnetic susceptibility introduced by
coupling of the magnetic field to the orbital current reveals an
anomalous temperature dependence, changing character from diamagnetic
to paramagnetic at intermediate temperatures.
\end{abstract}

\maketitle

\section{Introduction}

Anomalous normal-state properties of superconducting cuprates
\cite{imad} have stimulated intense theoretical investigations of
models of strongly correlated electrons describing the interplay
between antiferromagnetic (AFM) ordering of reference (undoped)
insulating substances and the itinerant character of charge carriers
introduced by doping. For the understanding of superconductivity the
most challenging regime is that of intermediate (optimum) doping.
However, even the apparently simplest region of weak doping is not
fully understood theoretically.

Recently, the attention in experimental and theoretical investigations
of cuprates has been given to characterization and understanding of
different doping regimes \cite{batl}. In a simple picture, weak doping
should correspond to the regime where properties vary linearly with
the concentration of holes, i.e.\ one can deal with a
semiconductor-like model where charge carriers (holes) are independent
and well defined quasiparticles. This requires a nonsingular variation
of thermodynamic quantities with doping. However, this scenario has
been questioned near the metal--insulator transition based also on
numerical solutions for some model systems \cite{imad}, e.g.\ the
Hubbard model. Alternative possibilities include phase separation
\cite{emer}, quantum critical behavior \cite{soko} or other
instabilities at low doping. Still, singular behavior in a planar (2D)
system is expected only at $T=0$, while $T>0$ should lead to a regular
variation with doping.

Among the least understood properties of charge carriers in cuprates
and correlated systems in general are those related to the coupling of
their orbital motion to an external magnetic field. Evidently
anomalous and not understood is the Hall constant in cuprates which
reveals unusual temperature and doping dependence \cite{ong}. Another
quantity is the diamagnetic (orbital) susceptibility $\chi_d$, which
for noninteracting electrons corresponds to Landau diamagnetism
\cite{land} and seems to be connected to the  Hall response
\cite{rojo}. Anomalous paramagnetic-like variation with magnetic field
has been noticed within the ground state of the $t$-$J$ model
\cite{bera} at low doping. Recent $T>0$ studies of a single hole
within the $t$-$J$ model \cite{vebe} confirm the existence of a
paramagnetic regime at intermediate $T$, though the systems studied were
quite small. Conclusive experimental results on diamagnetic
susceptibility are lacking \cite{wals}, since the orbital part appears
quite hidden by other contributions, although it could be distinguished via
the anisotropy.

The aim of this paper is to study the thermodynamic properties and
orbital response of correlated electrons at finite temperature in the
low-doping regime. Most numerical studies of the $t$-$J$ model have so
far focused on the ground-state properties \cite{dago}, employing
exact diagonalization of small systems, projector Monte Carlo, and
density matrix renormalization group \cite{whit} (DMRG). Recently, the
finite-temperature Lanczos method (FTLM) has been introduced, which
allows insight into the statics and dynamics at $T>0$. In previous
applications certain thermodynamic quantities have also been
investigated as a function of doping. In this paper we focus on the
low doping regime, where the method can be compared with the
alternative approach, a novel adaptation of the worldline quantum
Monte Carlo (QMC) cluster method \cite{ever} which allows for the
study of much larger systems at least for temperatures $T>T_-$ below
which the minus-sign problem sets in. Large systems are particularly
important for the study of diamagnetic response which appears to be
quite sensitive to finite size effects. In both cases, new ways of
dealing with the magnetic field are introduced. Related QMC methods
have been used to study nonmagnetic properties of the $t$-$J$ model,
in an exploratory calculation for doped chains and for ladders with 1
and 2 holes \cite{ammo}, in two dimensions at $J\rightarrow 0$ with 1
or 2 holes \cite{brun}, and for chains at finite $J$ in a background
of no holes \cite{assa}.

In the following, the planar $t$-$J$ model as a representative model
for strongly correlated electrons and electronic properties of
cuprates is studied,
\begin{equation}
H=-t\sum_{\langle ij\rangle\sigma}(\tilde{c}^\dagger_{j\sigma}
\tilde{c}^{\phantom{\dagger}}_{i\sigma}+\text{H.c.})+
J\sum_{\langle ij\rangle}\left(\vec{S}_i\cdot\vec{S}_j
-\tfrac{1}{4}n_in_j\right),
\label{eq:model}
\end{equation}
where $\tilde{c}^\dagger_{i\sigma}$,
$\tilde{c}^{\phantom{\dagger}}_{i\sigma}$ are fermionic operators,
projecting out sites with double occupancy. To approach the regime of
strong correlations close to the real situation in cuprates, $J/t=0.4$
is used in most numerical calculations. We also use $k_B=\hbar=1$.

The paper is organized as follows. Section II of the paper is devoted
to a brief introduction of both numerical techniques employed, QMC and
FTLM. In Sec.\ III results for several thermodynamic properties in the
low-doping regime are presented and discussed. Sec.\ IV is devoted to
the discussion of the orbital susceptibility of the system.

\section{Numerical Methods}

Results are obtained independently by the worldline QMC method and the
FTLM. Wherever possible, results of both methods for doped systems are
compared and presented relative to the undoped Heisenberg AFM. For
large enough systems we expect to reach a typical behavior in the low
doping regime.

\subsection{Worldline quantum Monte Carlo method}

The loop cluster algorithm (LCA) for the world-line QMC has been
introduced by one of the present authors \cite{ever} and recently
adapted also to the $t$-$J$ model \cite{ammo,brun}.

We briefly describe the worldline representation of the quantum
QMC. The Hamiltonian, Eq.~\ref{eq:model}, on a 2D square lattice can
be split within the standard Trotter-Suzuki decomposition
\cite{trot,suzu} into four parts $H=H_1+H_2+H_3+H_4$ consisting of
mutually commuting terms. This is equivalent to the well known
checkerboard decomposition of Hamiltonians in 1D. The partition
function is
\begin{eqnarray}
Z&=&\tr e^{-\beta H}=\lim_{M\to\infty}\tr[e^{-\tbeta(H_1+H_2+H_3+H_4)}]^M=
\nonumber\\
&=&\tr[e^{-\tbeta H_1}e^{-\tbeta H_2}e^{-\tbeta H_3}e^{-\tbeta H_4}]^M+
O(\tbeta^2)\approx
\nonumber\\
&\approx&\sum_{\phi_1\ldots \phi_{4M}}\bra{\phi_{4M}}e^{-\tbeta
H_1}\ket{\phi_1}
\bra{\phi_1}e^{-\tbeta H_2}\ket{\phi_2}\cdots
\nonumber\\
&\cdots&\bra{\phi_{4M-1}}e^{-\tbeta H_4}\ket{\phi_{4M}},
\label{eq:trotter}
\end{eqnarray}
where $\tbeta=\beta/M$ and $\beta=1/T$. The summation is taken over
the complete orthonormal set of states $\ket{\phi_i}$. Within each
imaginary time step $\tbeta$ the time evolution operator is
applied. Since the Hamiltonian is total spin conserving, we can track
time evolution of a particular spin along its so called {\em
worldline} (WL). Because of the cyclic property of the trace, the WLs
are periodic in the imaginary time interval $[0,\beta]$. The time
evolution operator acts only on $2\times2$ plaquettes and the weight
of the configuration $W(\conf)$ factorizes into a product of plaquette
weights. The partition function
\begin{equation}
Z=\sum_{\conf}W(\conf)=\sum_{\conf}\prod_{p\in\conf}W(p)
\label{eq:z}
\end{equation}
is formally that of a ($2+1$)-dimensional classical system. The
thermal average of an observable $\cal O$ can be obtained by
\begin{equation}
\avg{\cal O}=\frac{1}{Z}\sum_{\conf}W(\conf){\cal O}(\conf).
\label{eq:avg}
\end{equation}
Such thermal expectation values are calculated by means of Monte Carlo
(MC) importance sampling, where a sequence of configurations $\conf_i$
(Markov chain) is constructed, which obeys detailed balance and
reproduces the correct Boltzmann distribution $W(\conf)/Z$. Thermal
expectation values now become simple averages
\begin{equation}
\avg{\cal O}=\lim_{K\to\infty}\frac{1}{K}\sum_{i}^{K}
{\cal O}(\conf_i).
\label{eq:mc_avg}
\end{equation}
In practice, Monte Carlo runs are finite, $K<\infty$, leading to
statistical errors which can be calculated from the standard deviation
of partial data sets \cite{ever}.

In standard local algorithms an update from one configuration $\conf$
to another $\conf'$ in the Markov chain represents a small local
change of the WLs. Therefore, consecutive configurations are highly
correlated, which drastically increases the necessary number of Monte
Carlo steps.  Such difficulties are overcome in the LCA \cite{ever}
which introduces global (nonlocal) stochastic updates that effectively
reduce the correlations. In the LCA formulation also the continuous
time limit $\tbeta\to0$ can be taken \cite{bear} avoiding the second
order systematic error of Eq.~\ref{eq:trotter}. For certain
observables improved estimators can be easily constructed allowing a
potential reduction of statistical errors. For more details we refer
to the introductory paper \cite{ever}.

The LCA has recently been adapted to the $t$-$J$ model. The update
procedure is split into three substeps, allowing the application of
the standard LCA for the $S=1/2$ antiferromagnetic Heisenberg model or
for free fermions in all three cases. Within each substep, only
updates between two of the possible three states ($\uparrow$,
$\downarrow$, and hole $\circ$) are performed.  For the weights of
particular plaquettes and other technical details we refer to
\cite{ammo}.

In case of negative weights $W(\conf)<0$, their magnitude $|W(\conf)|$
is taken for construction of the MC procedure, since the negative
$W(\conf)$ cannot be taken as a probability.  Eq.~(\ref{eq:avg})
becomes
\begin{equation}
\avg{\cal O}=\frac{\avg{\text{sign}\cdot{\cal O}}_{|W|}}
{\avg{\sign}_{|W|}},
\label{eq:abs_avg}
\end{equation}
where $\avg{\cdots}_{|W|}$ denotes the expectation value with respect
to the absolute value of the weight.
In systems with such a ``sign problem'', the average sign
$\avg{\sign}_{|W|}$ often becomes exponentially small with increasing
system size and decreasing temperature $T$, leading to a blow up of
statistical errors \cite{suzu}.

Let us briefly comment on the origin of negative signs in the WL
formulation of the $t$-$J$ model. In the system with no doped holes
the only source of negative weights are plaquettes, where two opposite
spins exchange their positions representing a spin flip. Because of
the periodicity of WLs in time direction and the absence of holes spin
flips always occur in even numbers, producing no net negative sign. In
the pure Heisenberg model, this sign can also be transformed away by
rotation of spins on one sublattice, resulting in all-positive
plaquette weights \cite{ever}.

For one hole doped into the AFM one would naively not expect a sign
problem, e.g., in this case there is no sign in the exact
diagonalization approach. Examining the particle WLs surrounding the
hole WL one finds, however, that an exchange of two fermions can occur
when $t\ne 0$ and $J\ne 0$, producing an odd number of spin flips,
i.e.\ a negative sign, as can be seen schematically in a small
$2\times 2$ system,
\begin{equation}
\begin{bmatrix}
\circ&&\bar\uparrow\cr
&\circlearrowleft&\cr
\uparrow&\phantom{\leftrightsquigarrow}&\downarrow
\end{bmatrix}
\Rightarrow
\begin{bmatrix}
\circ&&\uparrow\\
&&\\
\downarrow&\leftrightsquigarrow&\bar\uparrow
\end{bmatrix}
\Rightarrow
\begin{bmatrix}
\circ&&\uparrow\\
&&\\
\bar\uparrow&\phantom{\leftrightsquigarrow}&\downarrow
\end{bmatrix},
\nonumber
\end{equation}
where a loop motion of the hole $\circlearrowleft$ around the system
and a consecutive spin flip $\leftrightsquigarrow$ reproduce the
original configuration with two fermions $\uparrow$ and $\bar\uparrow$
interchanged. Measuring the sign here reduces to spin flip counting.

For higher concentration of holes a more general expression for the
sign of the configuration can be obtained. It links fermion WL
($\text{perm}_f$) and hole WL permutation ($\text{perm}_h$)
\begin{equation}
\sign(\conf)=(-1)^{\text{perm}_f}=(-1)^{\text{perm}_h}\cdot
\sign(W(\conf)),
\end{equation}
so that for low doping it is preferable to measure $\text{perm}_h$
rather than $\text{perm}_f$. The sign problem also complicates the use
of the improved estimators since for every observable a separate
algorithm must be devised.

To follow the emergence of the sign problem as well as the development
of diamagnetic properties it is convenient to generalize the isotropic
spin interaction term of the model Eq.~(\ref{eq:model}) to an
anisotropic one with general anisotropy parameter $\gamma$,
\begin{equation}
H_J=J\sum_{\langle ij\rangle} [
\frac{\gamma}{2}(S_i^+S_j^-+S_j^+S_i^-)+S_i^zS_j^z].
\end{equation}
This modifies the pure spin substep ($\uparrow$, $\downarrow$) of the
$t$-$J$ LCA so that otherwise independent loops are ``frozen''
together \cite{ever} into clusters and updated stohasticaly.

\begin{figure}
\centering
\epsfig{file=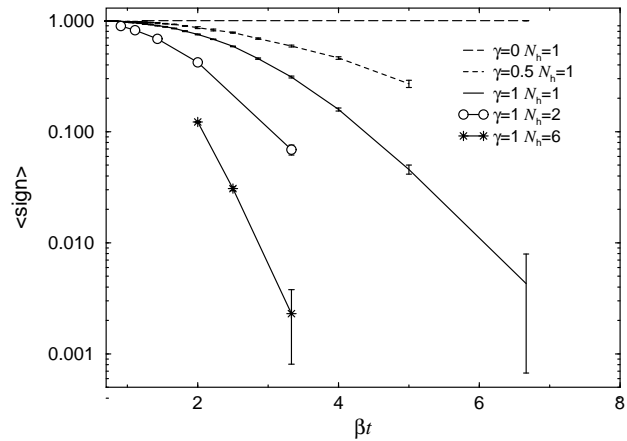,height=8.2cm,angle=-90}
\caption{Average sign in QMC for different values of the anisotropy
$\gamma$ and number of holes $N_h$ for a $6\times6$ system with
$J=0.4t$.}
\label{fig:sign}
\end{figure}

The results for $\avg{\sign}$ as a function of inverse temperature
$\beta t=1/k_BT$ are presented in Fig.~\ref{fig:sign}. Note that for a
single hole in the system, $N_h=1$, the relevant temperature scale in
the anisotropic case is $T_-\sim\gamma J$, i.e.\ there is no sign
problem for $\gamma=0$. At $T\lesssim T_-$ the sign starts to
deteriorate rapidly, as can be seen in Fig.~\ref{fig:sign}, preventing
the investigation of low temperature properties. As expected
$\avg{\sign}$ decreases by adding additional holes $N_h>1$. For this
reason, within the doped $t$-$J$ model only chains and coupled chains
\cite{ammo} have been investigated by LCA so far, and in two
dimensions the limit $J\to 0$ with $N_h=1,2$ \cite{brun}. Recently,
though, a way around the sign problem has been proposed by calculating
fermion propagators for a background of no holes \cite{assa}.

For a fixed number of holes, the average sign will converge as the
system size increases. This convergence could be taken as a another
criterion that the limit of a dilute system has been reached.

\subsection{Finite-temperature Lanczos method}

In the analysis of the $t$-$J$ model the exact diagonalization of
small systems using the Lanczos algorithm has been extensively
employed \cite{dago}, predominantly in the investigation of the static
and dynamic properties of the ground state. More recently a FTLM
combining the Lanczos procedure and random sampling was introduced
\cite{jakl1,jakl2}, allowing the calculation of $T>0$ static and dynamic
properties of correlated systems. The application is particularly
simple for an arbitrary function of conserved quantities,
e.g.\ $f(H,S_z)$,
\begin{eqnarray}
Z&\approx&\frac{N_{\text{st}}}{K}\sum_{n=1}^K\sum_{i=0}^{M-1}
e^{-\beta E_i^n}|\bracket{n}{\psi_i^n}|^2,
\label{eq:ftlm_z}
\\
\avg{f}&\approx&\frac{N_{\text{st}}}{KZ}\sum_{n=1}^K\sum_{i=0}^{M-1}
f(E_i^n,S_z^n)e^{-\beta E_i^n}|\bracket{n}{\psi_i^n}|^2,
\label{eq:ftlm_avg}
\end{eqnarray}
where $\ket{\psi_i^n}$, $E_i^n$ are (approximate) eigenfunctions and
energies, respectively, obtained by diagonalization within the reduced
orthonormal set, generated from the initial functions $\ket{n}$ in $M$
Lanczos steps. $N_{\text{st}}$ is the dimension of the complete
basis. $K$ initial functions $\ket{n}$ are chosen at random but with
good quantum number $S_z$. Usually it is enough to choose $M,K\ll
N_{st}$. For a more detailed discussion of the method and results we
refer to \cite{jakl2}.

It is expected that $T>0$ reduces the finite-size effects of the
measured quantities. It is however important to realize that for a
particular system, finite-size effects start to be pronounced at
$T<T_{\text{fs}}$ where, e.g., some characteristic length-scale
becomes larger than the system size. In our case of low doping,
$N\le20$ and $J=0.4 ~t$, we find $T_{\text{fs}}\sim 0.4J$. All our
results are presented for $T>T_{\text{fs}}$ where $Z(T_{\text{fs}})\sim
Z^*$. In the present study $Z^*=30$ so that at least 30 states are
sampled in the thermal averages \cite{jakl2}. It should be stressed
that the FTLM gives also the correct ground state within the chosen
small system.

\section{Thermodynamic Properties}

Thermodynamic properties ${\cal O}(c_h)$ depend on the hole
concentration $c_h=N_h/N$. For the weak doping limit, one would expect
a linear dependence for most quantities. For a finite system size, the
relevant parameter is thus the number of holes $N_h$ doped into the
AFM. In the low-doping regime it makes sense to represent the results
as a difference
\begin{equation}
\Delta{\cal O}_{i}={\cal O}(N_h\!=\!i)-{\cal O}(N_h\!=\!i-1).
\end{equation}
To distinguish the change in a particular quantity with doping, this
notation is used in the following. If, e.g., $\Delta{\cal O}_2$
behaves quantitatively as $\Delta{\cal O}_1$ one can conclude that the
quantity changes linearly with the number of added holes $N_h$, i.e.\
the holes behave as independent entities, and the system sizes are
large enough so that the low-doping regime has indeed been
reached. Such behavior is however not the only possibility at low
doping, since one can expect, e.g., even-odd effects in the case of
pairing of holes.

\subsection{Internal energy, specific heat and entropy}

\begin{figure}
\centering
\epsfig{file=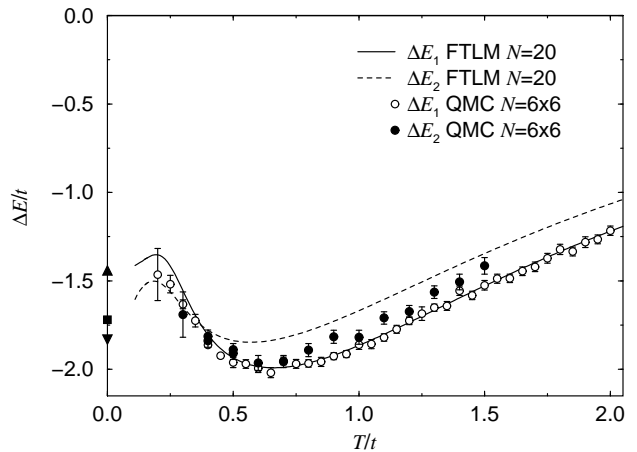,height=8.2cm,angle=-90}
\caption[]{Internal energy per hole $\Delta E$ vs.\ $T$ as calculated
within the FTLM and the QMC method for $J=0.4t$. $\blacksquare$
represents the one-hole ground state result for $N=4\times4$ from
Ref.~\cite{dago}. $\blacktriangle$ and $\blacktriangledown$ are our
ground state results for $\Delta E_1$ and $\Delta E_2$ for $N=20$,
respectively.}
\label{fig:eng}
\end{figure}

The internal energy, defined by
\begin{equation}
E=\frac{\partial\beta F}{\partial\beta}=-\frac{1}{Z}
\frac{\partial Z}{\partial\beta},
\label{eq:int_e}
\end{equation}
is calculated within FTLM as $\avg{E}$ in Eq.~(\ref{eq:ftlm_avg}) and
in QMC as an expectation value of the corresponding operator,
$E=\sum_{\conf}E(\conf)W(\conf)$. Results of both methods are
presented in Fig.~\ref{fig:eng}.

From Fig.~\ref{fig:eng} we first conclude that for $\Delta E_1$ the
results at $N=20$ (which overlap also with the results at $N=18$)
obtained via the FTLM are essentially equivalent with QMC results for
much larger lattices, at least for $T\geq0.2~t$ reached by the QMC
method. We note also a close agreement between QMC $\Delta E_2$ and
$\Delta E_1$, confirming the assumption of the low-doping regime and
holes as independent quasiparticles in the $T$ window presented. In
FTLM results, on the other hand, the difference between $\Delta E_2$
and $\Delta E_1$ is already visible, since $N_h=2$ here means already
an appreciable doping $c_h=0.1$. For $T=0$ the difference $\Delta
E_2-\Delta E_1$ equals to the binding energy \cite{dago} and in the
continuum corresponds to the second derivative of the ground state
energy with respect to the doping. In the chosen parameter regime
($J=0.4t$) the binding energy is negative and thus pointing to the
attractive interaction between the holes. With the increase of the
temperature the bound state disintegrates and the difference $\Delta
E_2-\Delta E_1$ approaches zero. In the case of a small system the
difference can even become positive but vanishes with increasing
system size.

It is also evident from Fig.~\ref{fig:eng} that $\Delta E(T)$ is not a
monotonous function. The ground state of a single hole introduced into
the AFM is quite well understood via analytical approaches
\cite{schm} and numerical calculations \cite{dago}. For $J/t=0.4$,
the zero temperature result $\Delta E(0)\sim-1.44t$ can be  explained well
by the interplay between the gain of the kinetic energy represented
by the hopping term $H_t$  and the loss of local AFM correlation energy
around the hole. 

$\Delta E(T)$ has not been considered so far. An interpretation of its
behavior can be given as follows. Introducing a single hole into an
AFM destroys the local AFM spin order and thus increases the exchange
energy. The increase is however expected to disappear at $T>J$ where
the spin system becomes disordered. On the other hand the ground-state
kinetic energy in a disordered spin system is quite similar to the one
in an AFM, hence the decrease of the internal energy $\Delta E$ for
$T>J$. This remains valid for $T<t$ where also higher hopping-related
states become populated and finally $\Delta E(T\to\infty)\to0$,
explaining turn back up for $T\gtrsim0.7t$.

\begin{figure}
\centering
\epsfig{file=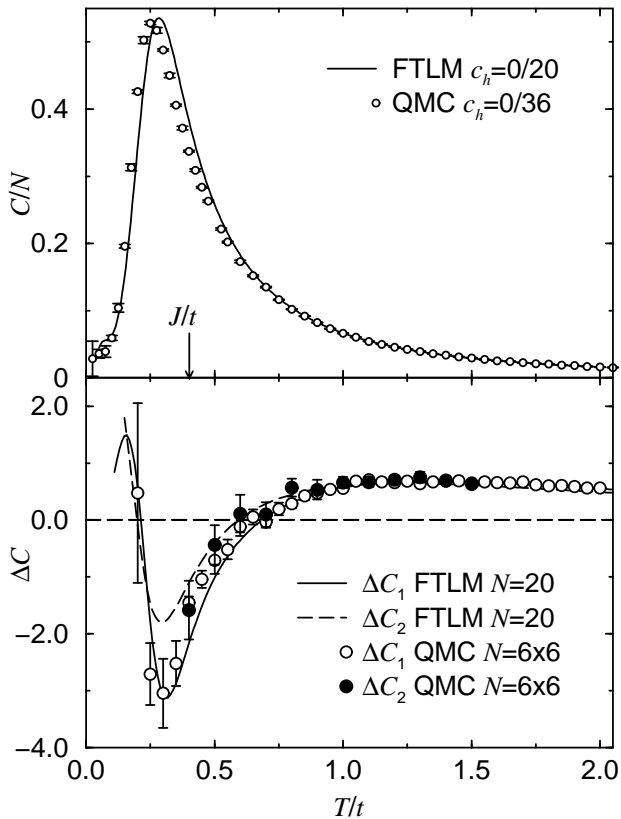,height=8.2cm,angle=-90}
\caption{Specific heat per site $C/N$ for undoped system (top) and change
with doping $\Delta C$ (bottom, both in units of $k_B$) vs.\ $T$ as
calculated within the FTLM and the QMC method for $J=0.4t$.}
\label{fig:heat}
\end{figure}

The specific heat defined by
\begin{equation}
C=\frac{\partial E}{\partial T}=
\beta^2\left[\frac{1}{Z}\frac{\partial^2Z}{\partial\beta^2}-
\left(\frac{1}{Z}\frac{\partial Z}{\partial\beta}\right)^2\right]
\end{equation}
is obtained as $\beta^2[\avg{E^2}-\avg{E}^2]$ within the FTLM and
the QMC method. The results are presented in Fig.~\ref{fig:heat}.

The main effect of introducing holes into the AFM insulator on $C$ is
to decrease the peak at $T\sim J$. This appears in $\Delta C$ as a
pronounced dip which slightly weakens and shifts its energy scale $J$
to lower values with doping, as can be seen from the line-shape in
Fig.~\ref{fig:heat}.

\begin{figure}
\centering
\epsfig{file=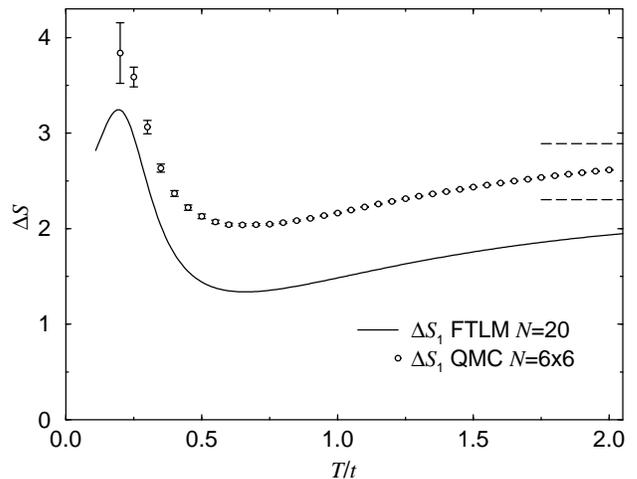,height=8.2cm,angle=-90}
\caption{Entropy increase $\Delta S$ (in units of $k_B$) for a single
hole obtained via the FTLM and the QMC method, on different size
lattices and $J=0.4t$. Long-dashed lines represent $\Delta
S(\infty)$.}
\label{fig:ent}
\end{figure}

The entropy is
\begin{equation}
S=\beta(E-F)=\beta E+\ln Z
\end{equation}
We reconstruct it from the specific heat
\begin{equation}
C=T\frac{\partial S}{\partial T}
\end{equation}
by numerical integration from high temperatures $T\sim\infty$
\begin{equation}
\Delta S(T)-\Delta S(\infty)=\int_\infty^T\frac{\Delta C}{T}\text{d}T
\end{equation}
The high-temperature integration constants are chosen so that
$\Delta S(\infty)=\Delta\ln N_{\text{st}}$.

In contrast to $\Delta E$ and $\Delta C$ discussed previously, $\Delta
S$ is not linear in $N_h$ in the low doping limit. In analogy with
low-concentration systems, like the dilute classical gas, entropy is
expected to scale as $\Delta S \propto |\Delta (N c_h \ln c_h)|$,
i.e. $\Delta S_1 \propto \ln N$, so the change still depends
explicitly on the system size. This is also realized in
Fig.\ref{fig:ent}, where $\Delta S_1$ still varies with $N$, yet
curves for different $N$ appear parallel down to the lowest reachable
temperatures.

It is particularly remarkable how large the entropy increase $\Delta
S_1 \gg 1$ is even at the lowest $T <J$. This is indeed consistent
with $\Delta S$ measured in cuprates \cite{lora}. While this could be
attributed partly to the logarithmic dependence on $c_h$, at the same
it is evident that the behavior is much closer to a system of
classical particles than to a degenerate electron gas.

\subsection{Spin susceptibility}

The uniform spin susceptibility can be evaluated as a thermodynamic
quantity from
\begin{equation}
\chi_s=\frac{\beta\avg{S_z^2}}{N},
\end{equation}
where $S_z=\sum_iS_i^z$ is the conserved total spin. In the FTLM then
Eq.~(\ref{eq:ftlm_avg}) can be applied, while within the QMC method
$\chi_s$ is related to the number of spin up and down WLs.

\begin{figure}
\centering
\epsfig{file=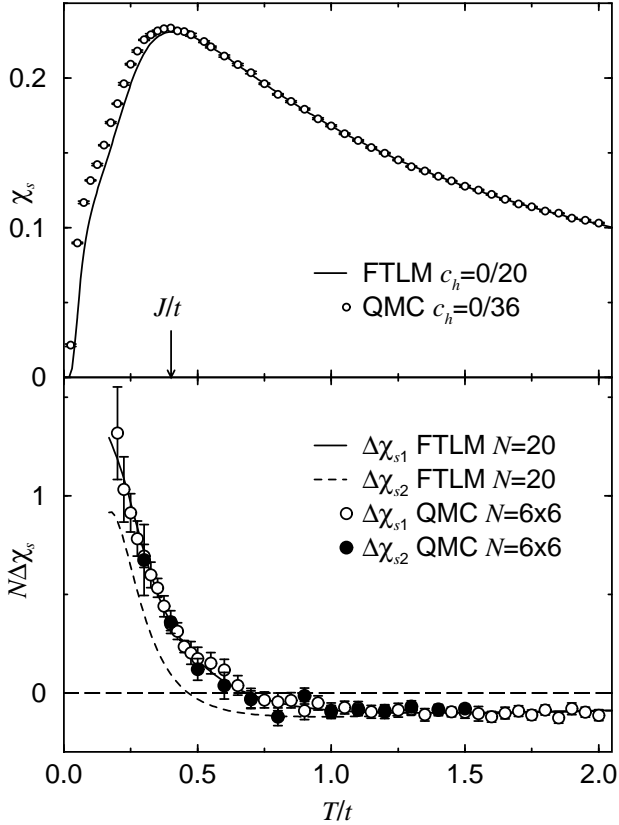,height=8.2cm,angle=-90}
\caption{Spin susceptibility for undoped system $\chi_s$ (top) and
change of the spin susceptibility with doping $\Delta\chi_s$ (bottom)
vs.\ $T$ obtained within the FTLM and the QMC method for $J=0.4t$.}
\label{fig:schi}
\end{figure}

\begin{figure}
\centering
\epsfig{file=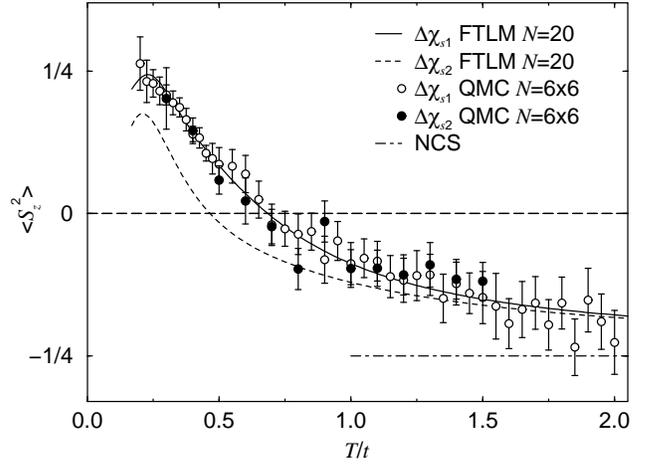,height=8.2cm,angle=-90}
\caption[]{Effective square moment (Curie constant) per hole
$\Delta\avg{S_z^2}$ vs.\ $T$, obtained from the data in
Fig.~\ref{fig:schi}. For comparison also the result for noninteracting
classical spins (NCS) is presented (dashed line).}
\label{fig:moment}
\end{figure}

It is instructive to present results both for $\Delta \chi_s$ with
respect to the undoped AFM, Fig.~\ref{fig:schi}, as well as for the
effective Curie constant (difference of the square moment) per hole
$\Delta\avg{S_z^2}=N\Delta\chi_s/\beta$ in Fig.~\ref{fig:moment}.

The results in Figs.~\ref{fig:schi}, \ref{fig:moment} are easy to
interpret for high $T>t$. Each hole introduced into the system reduces
the effective Curie constant by one spin, i.e.,
$\Delta\avg{S_z^2}=-1/4$. On the other hand, at low $T<J$ the
situation is reversed since $\Delta\chi_s>0$. This increase can be
attributed to the relaxation of the AFM order by the hole doping. Note
that in an AFM, $\chi_s$ achieves a maximum at $T\sim J$ while below
that temperature it is reduced due to the longer range AFM order. It
is interesting to note that at the lowest reachable temperature $T\sim
J/2$ each hole effectively adds just one spin, i.e.\
$\Delta\avg{S_z^2}\sim1/4$.

\section{Orbital Susceptibility}

In order to investigate the orbital response of the system, a
homogeneous magnetic field $B$ perpendicular to the plane has to be
introduced. When we discuss the orbital magnetization and
susceptibility, $B$ enters only in the kinetic term of
Eq.~(\ref{eq:model}), via the Peierls construction
\begin{equation}
H_t=-t\sum_{\langle ij\rangle\sigma}(e^{i\theta_{ij}}
\tilde{c}^\dagger_{j\sigma}
\tilde{c}^{\phantom{\dagger}}_{i\sigma}+\text{H.c.}),
\label{eq:peierls}
\end{equation}
where the phases are given within Landau gauge as
\begin{equation}
\theta_{ij}=\frac{e}{\hbar}\vc{A}(\vc{r}_i)\!\cdot\!\vc{R}_{ij},\qquad
\vc{A}=B(0,x,0),
\label{eq:theta}
\end{equation}
with $\vc{R}_{ij}=\vc{r}_j-\vc{r}_i$. The relevant parameter for the
strength of $B$ is the dimensionless flux per plaquette $\alpha=2\pi
Ba^2/\phi_0$, where $a$ is the lattice spacing and $\phi_0=h/e$ is the
unit quantum flux.

The dc orbital susceptibility of the system in the external
magnetic field is
\begin{eqnarray}
\chi_d&=&-\mu_0\frac{\partial^2 F}{\partial B^2}=
-\frac{\mu_0e^2a^4}{\hbar^2}\frac{\partial^2 F}{\partial\alpha^2}=
\nonumber\\
&=&-\frac{\chi_0}{\beta}\left[\frac{1}{Z}
\frac{\partial^2 Z}{\partial\alpha^2}-
\left(\frac{1}{Z}\frac{\partial Z}{\partial\alpha}\right)^2\right],
\label{eq:chidef}
\end{eqnarray}
where $\chi_0=\mu_0e^2a^4/\hbar^2$. So far $\chi_d$ has been
investigated only for a single hole $N_h=1$ by high-temperature
expansion at $J=0$, and by the FTLM for $J>0$ \cite{vebe}. In the
latter study it was realized that results are quite sensitive to
finite-size effects, so it is desirable to get corresponding results
also via the QMC method, where much larger lattices can be studied.

Let us first derive the expression for the orbital susceptibility
within the QMC method. As seen from Eq.~(\ref{eq:peierls}), the
magnetic field affects only the hopping of the electrons. In the WL
representation this concerns matrix elements $\bra{\phi}e^{-\tbeta
H}\ket{\phi'}$ in Eq.~(\ref{eq:trotter}). For the plaquette
representing the hopping between sites $i$ and $j$ the matrix elements
become
\begin{equation}
H_{ij}^{\text{hop}}=
\begin{pmatrix}
0&-te^{i\theta_{ij}}\\
-te^{-i\theta_{ij}}&0
\end{pmatrix},
\end{equation}
written in the $\ket{\circ\uparrow}$, $\ket{\uparrow\circ}$ base. The
imaginary time propagator in the same base is
\begin{equation}
e^{-\tbeta H_{ij}^{\text{hop}}}=
\begin{pmatrix}
\ch\tbeta t&e^{i\theta_{ij}}\sh\tbeta t\\
e^{-i\theta_{ij}}\sh\tbeta t&\ch\tbeta t
\end{pmatrix}.
\end{equation}
Thus the plaquette weights along the hole WL obtain an additional
phase factor $W(p,\alpha)=W(p)e^{i\theta(p)}$. The weight of the whole
configuration is a product, Eq.~(\ref{eq:z}),
\begin{equation}
W(\conf,\alpha)=\prod_{p\in\conf}W(p,\alpha)=W(\conf)
\prod_{p\in\conf}\exp\left(i\theta(p)\right)
\end{equation}
where the phases sum up
\begin{align}
\exp\Big[i\sum_{p\in\conf}\theta(p)\Big]=
\exp\Big[i\oint\theta(\vc{r})\text{d}\vc{r}\Big]=
\nonumber\\
=\exp\Big[i\frac{e}{\hbar}\oint\vc{A}(\vc{r})\text{d}\vc{r}
\Big]=e^{i\alpha{\cal S}}.
\label{eq:phases}
\end{align}
The integral runs along the hole WL. Here $\cal S$ is defined as the
oriented area of the hole WL projected onto the plane in units of the
lattice plaquette area $a^2$. For more holes, $\cal S$ generalizes
similarly to the sum of all hole WL areas.

Now we can write the partition function in the magnetic field as
\begin{equation}
Z=\sum_{\conf}W(\conf,\alpha)=
\sum_{\conf}e^{i\alpha{\cal S}(\conf)}W(\conf).
\label{eq:zz}
\end{equation}
For a given configuration $\cal C$ there always exist $\cal C'$
(imaginary time inversion) with the same weight but ${\cal S}({\cal
C'})=-{\cal S}({\cal C})$, therefore the exponential in
Eq.~(\ref{eq:zz}) can be replaced by a $\cos$ function. For $B=0$ we
have $\avg{\cal S}=0$ and obtain the zero-field susceptibility from
Eq.~(\ref{eq:chidef}),
\begin{equation}
\chi_d=-\chi_0\frac{\avg{{\cal S}^2}}{\beta}.
\label{eq:chi}
\end{equation}
$\chi_d$ can be thus measured without the presence of a magnetic
field.  This is just another consequence of the more general
fluctuation--dissipation theorem. Even though ${\cal S}^2$ is strictly
positive, the thermal average in Eq.~(\ref{eq:chi}) can become
negative because of correlations between the sign of the weight and
the area $\cal S$ of the hole WL. From Eq.~(\ref{eq:abs_avg}) we can
deduce that $\avg{{\cal S}^2}<0$ when the configurations with negative
sign tend to have larger ${\cal S}^2(\conf)|W(\conf)|$ than
configurations with positive ones.

The hole WL can obtain nonzero spatial winding number due to the
periodic boundary conditions and small system size. E.g., the WL can
run along the imaginary time, cross the system boundary, and complete
time periodicity reconnecting with its spatially periodic image. In
that case the area $\cal S$ as defined by Eq.~(\ref{eq:phases}) has no
physical meaning. Therefore we restrict our simulation so that only
the zero spatial winding loop updates are generated (a discussion on
fixing the winding numbers can be found in \cite{hene}). The hole is
allowed to cross the system boundary as long as it does not increase
the winding number. The effect of the restriction is analogous to the
movement of the hole doped into an infinite periodic spin background
with a unit cell equal to the size of the system. The results for
thermodynamic quantities presented in the previous section agree
within the errorbars with the unrestricted case. This restriction is
weaker than closed boundary conditions, resulting in smaller finite
size effects.

\begin{figure}
\centering
\epsfig{file=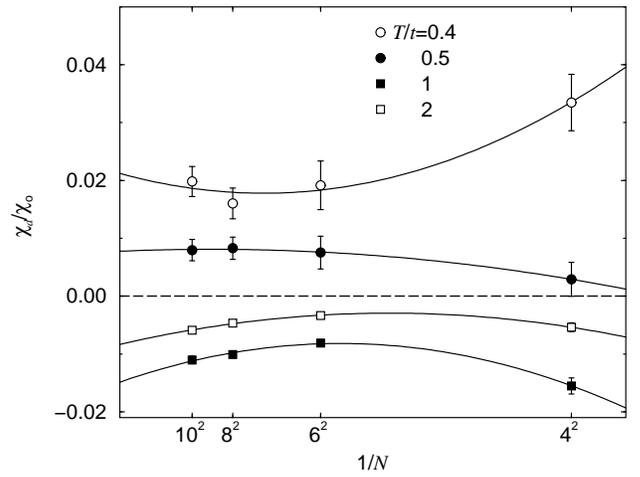,height=8.2cm,angle=-90}
\caption{Finite-size scaling of $\chi_d$ from QMC with one hole, for
different $T$ and $J=0.4t$. Solid lines serve as a guide to the eye
and are fits to the $1/N^2$ and $1/N^4$ dependence.}
\label{fig:scale}
\end{figure}

The introduction of finite $B>0$ into the model,
Eq.~(\ref{eq:peierls}), reduces the translational symmetry and thus
for a given system size increases the required minimal base set used
in FTLM. In the present study a few mobile holes on a system of tilted
squares with $N$ up to 20 sites and periodic boundary conditions are
considered. It is nontrivial to incorporate phases $\theta_{ij}$
corresponding to a homogeneous $B$, being at the same time compatible
with periodic boundary conditions \cite{frad,vebe}. This is possible
only for quantized magnetic fields $B=m B_0$, where $B_0=\phi_0/N$.

$\chi_d$ from Eq.~(\ref{eq:chi}) can be calculated in FTLM only by
taking a numerical derivative of the free energy $F=-T\ln Z$, with $Z$
from Eq.~(\ref{eq:ftlm_z}). Finite systems provide $F(\alpha)$ only
for discrete values of $\alpha$. $\chi_d$ can be obtained by fitting
the $F(\alpha)$ dependence to the parabolic form
$F(\alpha)=F(0)+c\alpha^2$ in two points $\alpha=i\cdot2\pi/N$ and
$\alpha=j\cdot2\pi/N$. The corresponding results for the
susceptibility are denoted by $\chi_{ij}$. In small systems ($N<20$)
the introduction of $B>0$ can lift some zero-field degeneracies. The
$\chi_{01}$ values are therefore systematically affected by larger
finite-size effects. For the system with $N=20$ both values
$\chi_{01}$ and $\chi_{12}$ agree quite reasonably with the QMC data.

Let us first comment on the validity of results for $\chi_d$. Since
$\chi_d$ deals with an orbital current represented by loop motion of
charge carriers (holes), it is much more sensitive to finite-size
effects \cite{vebe} than most other correlation functions. This was
also the main motivation to employ the QMC method, where much larger
systems can be reached. In Fig.~\ref{fig:scale}, finite-size scaling
is performed for the QMC data for $\chi_d(T)$ for the case of
$N_h=1$. We can see that the different $T$ points do not cross upon
changing the system size $N$. Thus, at least qualitatively, results do
not depend on the system size. Therefore, the sizes of choice for the
QMC systems will be $6\times6$ and $8\times8$.

\begin{figure}
\centering
\epsfig{file=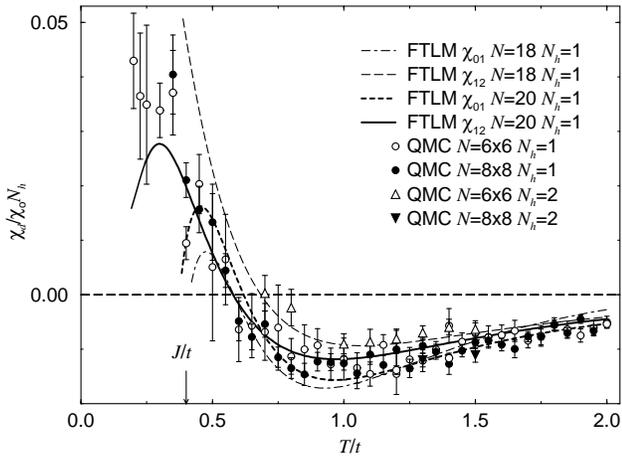,height=8.2cm,angle=-90}
\caption{Orbital susceptibility $\chi_d$ vs.\ $T$ obtained via the QMC
method and the FTLM method for $J=0.4t$. Regarding the FTLM
$\chi_{12}$ for $N=20$ should be most relevant (thick line).}
\label{fig:dchi}
\end{figure}

\begin{figure}
\centering
\epsfig{file=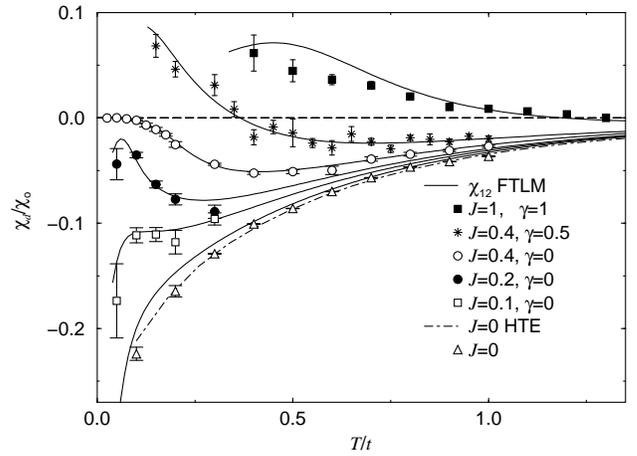,height=8.2cm,angle=-90}
\caption[]{$\chi_d$ vs.\ $T$ for $N_h=1$ and different $J$ and $\gamma$.
QMC results for the $6\times 6$ system are shown with different dots
and are labeled with appropriate values of $J$ and $\gamma$. Solid
lines correspond to the $N=20$ FTLM for the same values of $J$,
$\gamma$ as in QMC. For comparison also results of high-temperature
expansion (HTE) for $J=0$ \cite{vebe} are shown (dash-dotted
line). The data for $J=0.4$, $\gamma=1$ from Fig.~\ref{fig:dchi} are
not shown again.}
\label{fig:gammaj}
\end{figure}

In Fig.~\ref{fig:dchi}, $\chi_d$ obtained via both methods is
presented. For $T\gg t$, the response is diamagnetic and proportional
to $T^{-3}$ as well as essentially $J$-independent \cite{vebe}. The
most striking effect is that the orbital response below some
temperature $T_p$ turns from diamagnetic to paramagnetic, consistent
with the preliminary results obtained via the FTLM \cite{vebe}. In
order to locate the origin of this phenomenon, results for different
$J$ and anisotropies $\gamma$ are shown in Fig.~\ref{fig:gammaj}. It
appears that $T_p$ scales with $\gamma J$, i.e.\ at $J=0$ the response
is clearly diamagnetic at all $T$, and for $\gamma=0, J>0$ no crossing
is observed with either method.

At lower temperatures $T<T_d\ll T_p$, the diamagnetic behavior is
expected to be restored. This follows from the argument that at
$T\to0$ a hole in an AFM should behave as a quasiparticle with a
finite effective mass, exhibiting a cyclotron motion in $B\neq0$. The
latter behavior should lead to $\chi_d(T\to0)\to-\infty$ \cite{vebe}.
Numerically it is easiest to test this conjecture for a single hole
and $\gamma=0$ (also true for $J=0$). Namely, the QMC has no sign
problem at $\gamma=0$, so that error bars are only due to the finite
MC sampling.

Results in Fig.~\ref{fig:gammaj} are quite interesting even for
$\gamma=0$. At $J=0$ a monotonous increase of $|\chi_d|$ is observed,
diverging as $T\to 0$ \cite{vebe}. It can be explained as a gradual
transition from a hole in a random spin background to a well defined
quasiparticle, i.e.\ the ferromagnetic polaron, at $T\to 0$. The
situation is more complicated for $J>0$. It is plausible that the
difference to the $J=0$ case shows up at $T<J$, where the AFM
short-range correlations appear. Relative to $J=0$, a spin ordered
state blocks the loop motion of holes, necessary for finite
diamagnetic $\chi_d$. The effect is thus first a decrease of
$|\chi_d(T)|$ with decreasing $T$, as seen in Fig.~\ref{fig:gammaj}.
Turnover to the diverging diamagnetic $\chi_d$ should happen only when
the coherent quasiparticle is formed at $T_d\ll J$. $T_d$ should scale
with the inverse of the quasiparticle effective mass $1/m^*$. It is
known that $m^*$ can become very large for the extreme $\gamma=0$
case, in particular for larger $J$. This explains why we cannot reach
the coherent regime for $J=0.4~t$ even at $T=0.2t$
(Fig.~\ref{fig:dchi}), while the downturn is indeed observed for
$J/t=0.1$, $0.2$, $\gamma=0$ (Fig.~\ref{fig:gammaj}).

At $\gamma >0$, the results are qualitatively different. The most
pronounced effect is the change into a paramagnetic $\chi_d$ for
$T<T_p$. The width of this $T$ window is quite large. In fact within
the FTLM and QMC data we are unable to locate the reentrance
temperature $T_d$ into the diamagnetic response, although the latter
is expected \cite{vebe}. An argument for the low value of $T_d$ can be
given in terms of a very shallow energy minimum which defines the
quasiparticle dispersion near the ground-state of a hole in an AFM
within the $t$-$J$ model \cite{dago}, hence the quasiparticle looses
its character already at very low excitation energies. Still the
paramagnetic response in the window $T_d<T<T_p$ remains to be
explained.

\begin{figure}
\centering
\epsfig{file=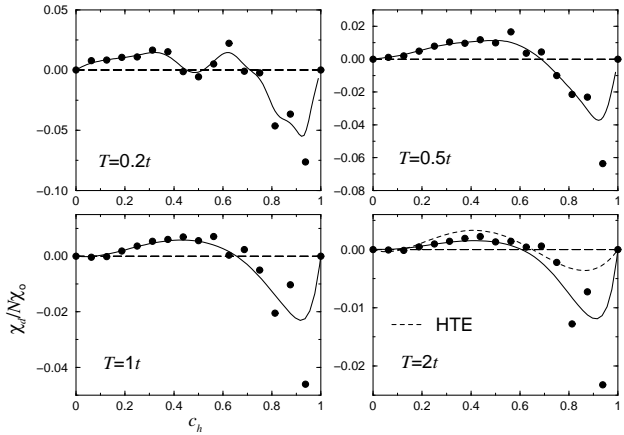,height=8.2cm,angle=-90}
\caption{$\chi_d$ vs.\ $c_h$ for several $T$ and $J=0.4t$. The last graph
contains also a 4$^{\text{th}}$ order HTE result (dotted). Canonical
(dots) and grand-canonical (line) values for all $c_h$ are obtained
within the FTLM with $N=16$, where $\chi_{12}$ is used to calculate
$\chi_d$.}
\label{fig:chich}
\end{figure}

Let us finally discuss also results for $\chi_d$ for finite doping
$c_h>0$, as presented in Fig.~\ref{fig:chich}. The easiest regime to
interpret is that of a nearly empty band, i.e.\ $c_h>0.7$, where
$\chi_d$ is diamagnetic and nearly independent of $T$. In this regime
the electron system is dilute and strong correlations are unimportant,
hence Landau diamagnetism is expected. At moderate temperatures $T>J$
and for an intermediate-doping regime, $0.2<c_h<0.7$, $\chi_d$ is
dominated by a paramagnetic response with a peak at approximately
$c_h=1/2$. As consistent with results at low doping, there is a weak
diamagnetism at $c_h<0.2$ and $T>T_p$, while the paramagnetic regime
extends to $c_h=0$ for $T<T_p$. For low temperatures $T \ll J$ quite
pronounced oscillations in $\chi_d(c_h)$ appear and can be partly
attributed to finite-system effects.

Certain aspects of the above results for $\chi_d(c_h)$ can be
understood using the HTE. One is the asymmetry between $c_h \to 0$ and
$c_h \to 1$ at higher $T>t$. In the lowest order of the HTE, only
hopping of the electrons around a basic plaquette loop has to be
considered. The signal on the $c_h\to0$ side of $\chi_d(c_h)$ is thus
reduced by a factor of $2^3=8$ against the $c_h\to1$ side since in the
first case only the plaquettes with ferromagnetically aligned spins
contribute.

\section{Conclusions}

Let us first compare both numerical methods used in the analysis of
the $t$-$J$ model at low doping. The QMC method allows the studies of
larger systems and the loop algorithm solves some serious drawbacks of
the MC methods. It is indeed very efficient in cases where there is no
sign problem, e.g.\ the anisotropic model $\gamma=0$ at $N_h=1$.
Within the WL approach it is also very easy to formulate and measure
certain responses like the orbital susceptibility $\chi_d$. Still the
method suffers from a sign problem (for $\gamma>0$) even for a single
hole $N_h=1$ in an AFM (though not in a background of no holes
\cite{assa}). Results are thus in practice limited to $T\gtrsim J/3$
for the isotropic case $\gamma=1$. On the other hand the FTLM has no
minus-sign problem but rather limitations due to small systems which
can be studied. These are even more pronounced in cases with $B>0$
where the translational symmetry is lost. It is an interesting
observation that within the FTLM, the limiting temperature $T_{fs}$
for the $t$-$J$ model is in most cases quite close to the lowest $T$
reached by the QMC method.

We have presented several results for thermodynamic quantities, i.e.\
energy $E$, specific heat $C$, entropy $S$ and spin susceptibility
$\chi_s$, as a function of $T$ at low doping. At $T>T_{fs}$ all
results are consistent with the picture where holes introduced into
the AFM behave as independent (nondegenerate) particles.
The perturbation introduced into the AFM by holes is quite large even
at lowest $T<J$, in particular visible from $\Delta C$ and $\Delta S$,
consistent with experiments in cuprates \cite{lora}.

Results for the orbital susceptibility $\chi_d$, now obtained also for
much larger systems using the QMC method, confirm the preliminary FTLM
results \cite{vebe}, indicating an anomalous paramagnetic response at
low doping in an intermediate window of temperatures $T_d<T<T_p\sim J$
(for the isotropic model $\gamma=1$). In fact within our numerical
studies it is hard to reach the lower end of this window, meaning that
$T_d<J/3$. Still, the reentrance into a diamagnetic behavior is
expected from theoretical arguments on the existence of a well defined
quasiparticle at $T\to 0$, as well as from more reliable QMC results
for the $\gamma=0$ case \cite{vebe}. The paramagnetic response at
intermediate $T$ can be viewed also as an extension of a more
pronounced $\chi_d>0$ regime observed at finite doping $0.2<c_h<0.5$
at all $T$. The explanation can thus go in the direction proposed by
Laughlin \cite{laug,bera}, that at low doping $c_h \to 0$ we are
dealing with quasiparticles (with a diamagnetic response), being a
bound composite of charge (holon) and spin (spinon) elementary
excitations. The binding appears however to be quite weak and thus
easily destroyed by finite $T$ or $c_h$, enabling the independent
response of constituents, which apparently is paramagnetic.

\begin{acknowledgements}

The authors wish to thank I.\ Sega for helpful suggestions. This work
was supported by the Ministry of Science and Technology of Slovenia
under Project No.~J1-0231.

\end{acknowledgements}

\end{document}